\documentclass[a4paper]{article}

\usepackage{INTERSPEECH2019}

\usepackage{booktabs}
\usepackage{threeparttable}
\usepackage{multirow}
\usepackage[colorlinks,
            linkcolor=red,
            anchorcolor=blue,
            citecolor=green,
            urlcolor=red,
            ]{hyperref}
\usepackage{float}

\newcommand{\xt}[1]{\textit{#1}}
\def\etal{\textit{et al. }}
\def\babycry{\textit{babycry}}
\def\glassbreak{\textit{glassbreak}}
\def\gunshot{\textit{gunshot}}
\def\mb{\textit{mask branch}}
\def\fb{\textit{feature branch}}

\title{Multi-Scale Time-Frequency Attention for Acoustic Event Detection}
\name{Jingyang Zhang$^1$, Wenhao Ding$^1$, Jintao Kang$^{2,3}$, Liang He$^1$\thanks{Liang He is the corresponding author.}} 
\address{
  $^1$Department of Electronic Engineering, Tsinghua University, China\\
  $^2$Institute of Forensic Science, Ministry of Public Security, China\\
  $^3$Key Laboratory of Intelligent Speech Technology, Ministry of Public Security, China}
\email{jz288@duke.edu, dingwenhao95@gmail.com, kangjintao@cifs.gov.cn, heliang@mail.tsinghua.edu}

\begin{document}

\maketitle
%
\begin{abstract}
  Most attention-based methods only concentrate along the time axis, which is insufficient for Acoustic Event Detection (AED). Meanwhile, previous methods for AED rarely considered that target events possess distinct temporal and frequential scales. In this work, we propose a \xt{Multi-Scale Time-Frequency Attention} (MTFA) module for AED. MTFA gathers information at multiple resolutions to generate a time-frequency attention mask which tells the model where to focus along both time and frequency axis. With MTFA, the model could capture the characteristics of target events with different scales. We demonstrate the proposed method on Task 2 of Detection and Classification of Acoustic Scenes and Events (DCASE) 2017 Challenge. Our method achieves competitive results on both development dataset and evaluation dataset.


\end{abstract}

\section{Introduction}

Acoustic Event Detection (AED) defines a task which requires identification of whether some certain events occur in an audio clip, and their timestamps should be given if they do exist. AED has drawn great attention among researchers since it is crucial for a system especially when visual cues are not sufficient or not available. For example, an autonomous car should be able to detect sirens to give way for an ambulance approaching from behind \cite{schroder2013automatic}. Household robots \cite{wu2009intelligent} and road surveillance systems \cite{foggia2016audio} can also benefit from AED.

To facilitate the research of AED, Task 2 of Detection and Classification of Acoustic Scenes and Events (DCASE) 2017 Challenge \cite{DCASE2017challenge} provides datasets and a well-designed testbench. The task asks participants to locate the onset of three classes of rare events (\babycry, \glassbreak, \gunshot) within audio clips.

One straightforward approach to this is making frame-level predictions, which was adopted by many participants including 1$^{\text{st}}$ and 2$^{\text{nd}}$ place \cite{Lim2017, Cakir2017} of DCASE2017 Task2. Kao \etal \cite{Kao2018}, on the contrary, applied \xt{Region Proposal Network} \cite{ren2015faster} to perform event-level detection and directly generate the onset/offset of target events. However, this method needs pre-training and suffers from over-fitting due to a large number of parameters. Wang \etal \cite{Wang2018} further incorporated attention mechanism to derive utterance-level predictions from frame-level output, attempting to leverage attention to improve frame-wise detection.

\begin{figure}[!t]
  \centering
  \includegraphics[width=3in]{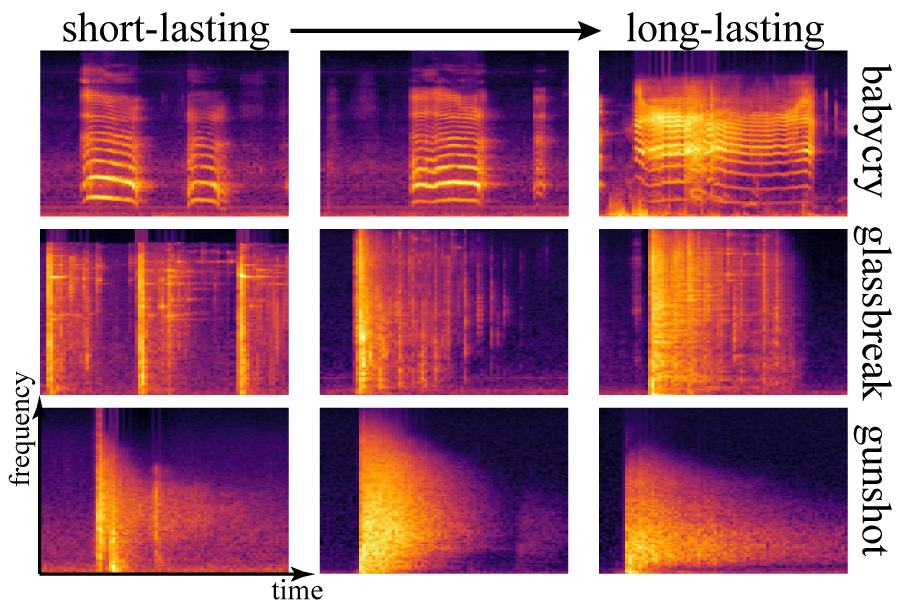}
  \caption{\textbf{Mel-spectrogram} of the events in DCASE2017 Task2.}
  \label{fig:mel}
\end{figure}

Following the majority of previous works, we produce frame-level predictions in this work. As attention mechanism has achieved excellent performance in various tasks like emotion recognition \cite{Luo2018}, conversational analysis \cite{Bothe2018}, and speech recognition \cite{Chang2018}, we assume it can also benefit AED as it helps the model focus on when target events take place. However, \cite{Wang2018} and most attention-based methods operate only along the time axis, which is insufficient for AED because events present different frequential characteristics. It can be seen in Figure \ref{fig:mel} that while \babycry{} and \glassbreak{} cover a wider range in the frequency domain, \gunshot{} consists of high-frequency components only at the beginning and soon low-frequency energies become dominant. This variation makes it necessary to tell the model where to focus along the frequency axis.

Besides, previous methods did not take into account that target events possess different scales. As shown in Figure \ref{fig:mel}, the duration or speed varies among not only different types of events, but also different samples within each class. Meanwhile, frequential scales might also change. For example, different types of guns present contrasting characteristics. Thus, if the model perform on a single resolution, it might miss useful information to capture the distinct properties of each event.

From the above two perspectives, inspired by \xt{Residual Attention} \cite{wang2017residual}, we propose a \xt{Multi-Scale Time-Frequency Attention} (MTFA) module for AED. MTFA fuses multi-scale information to generate a 2D time-frequency attention heatmap and produces attention-aware representations. The generated attention weights instruct the model to focus on not only important frames along the time axis but also import components along the frequency axis. To sum up, MTFA is designed to extract powerful features by leveraging multi-scale learning and attention mechanism. To the best of our knowledge, the proposed method outperforms previous single-model methods on development dataset of DCASE2017 Task2 and achieves state-of-the-art on evaluation dataset by reducing the error rate to 0.09 from previously best 0.13. The result provides clear evidence that MTFA has a strong ability to obtain discriminative representations.

\begin{figure*}[!t]
  \centering
  \includegraphics[width=6.5in]{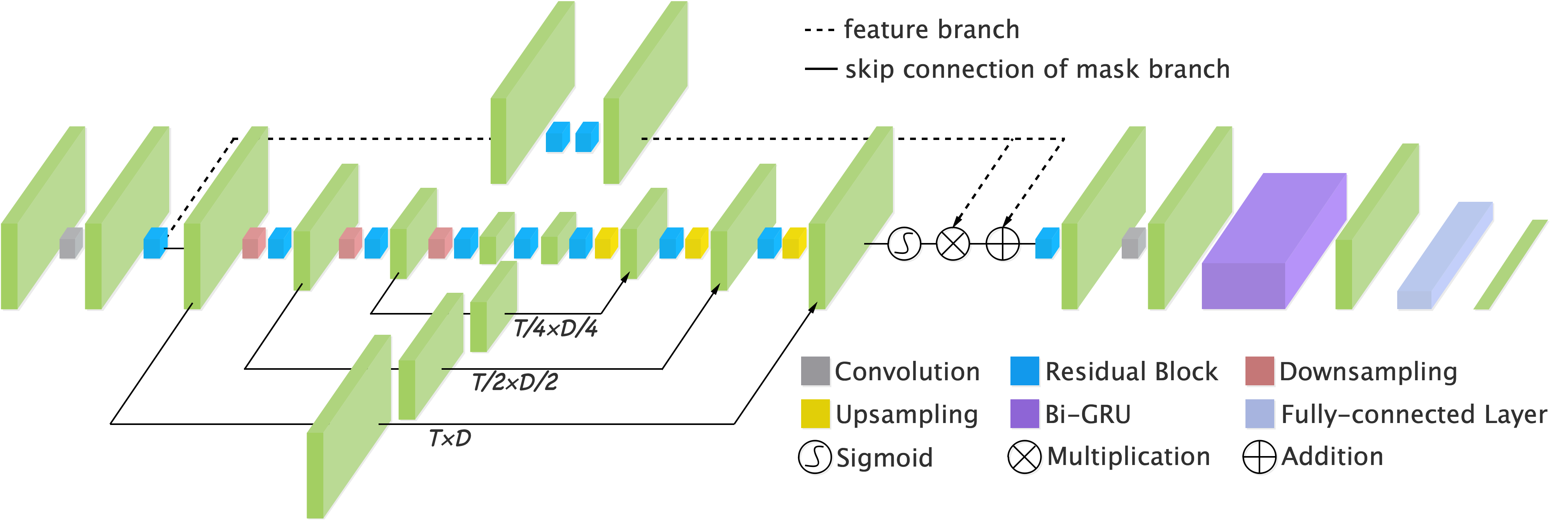}
  \caption{\textbf{Global structure of the proposed method.} Each green block represents a feature map or the raw input. MTFA consists of a \textbf{feature branch} for feature extraction and a \textbf{mask branch} for attention generation. The output of two branches is combined in a residual manner to produce multi-scale attention-aware features. A bi-directional GRU is applied to capture temporal dependencies between frames on top of MTFA. Finally, fully-connected layers will give frame-wise predictions.}
  \label{fig:topology}
\end{figure*}

\section{Method Description}

We present the overall model structure in Figure \ref{fig:topology}. The network mainly consists of three modules: a \xt{Multi-Scale Time-Frequency Attention} module, a Recurrent Neural Network, and a classifier. The MTFA extracts multi-scale attention-aware features with a focus on important frames and frequential components. The RNN further captures temporal information. The classifier produces frame-level predictions which indicate whether the frame corresponds to the event.

\subsection{Input Acoustic Features}
Similar to previous works \cite{Lim2017, Cakir2017, Kao2018, Wang2018, shen2018learning}, we use log-scale filter banks to obtain Mel-spectrogram as the raw input. Specifically, we first apply a 40ms sliding window with a 20ms shift over the audio signal and then calculate 128-dimensional log filter bank energies for each frame. Finally, we will get a 2D time-frequency representation with a size of $(1501\times128)$ for each 30-second sound clip.

\subsection{Multi-Scale Time-Frequency Attention}
The intention of MTFA are two folds: \xt{(1)} gathering information at multiple resolutions to model different scale characteristics presented by target events and \xt{(2)} producing both temporal and frequential attention weights to focus on not only important frames but also discriminative frequency components. Note that these two ideas are not separate or independant.

First, we believe the fusion of multi-scale information is beneficial for AED. As reported by \cite{shen2018learning}, the average duration of the events are: 2.25s for \babycry{}, 1.16s for \glassbreak{}, and 1.32s for \gunshot{}. This is quite intuitive as the latter two events mainly consist of short-time impulses, and \babycry{} generally lasts for a longer time. If we observe on one single time resolution, it might be insufficient to model either the long-lasting \babycry{} or the relatively transient \glassbreak{} and \gunshot{}. \cite{Wang2018} already extracted multi-resolution features along the time axis to improve the robustness against this time variation. However, frequency variation also exists in AED, as demonstrated by Figure \ref{fig:mel}. Due to the limitation of human sounds, low-frequency components are the major part of \babycry{}, but \glassbreak{} has nearly every frequency ingredient. \gunshot{} differs a lot from them as it starts with high-frequency impulses and vanishes with low-frequency energies. The change of gun types may also results in distinct properties. Therefore, in this work, we design MTFA to extract representations from multiple scales in both the time and frequency domain.

Second, to help the model focus more on the timestamp of target events and meanwhile pick out discriminative frequency components, we propose to generate both temporal and frequential attention. Specifically, by utilizing convolution as the basic operator, our method takes in a spectrogram and generates a 2D time-frequency attention heatmap, which contains element-wise attention weights. In other words, MTFA leverages the 2D nature of convolution and time-frequency features, thus enabling itself to take care of temporal and frequential attention simultaneously. Besides, the multi-channel nature of convolution provides the model with the capacity of collecting richer information to generate attention.

Specifically, our MTFA module is inspired by \xt{Residual Attention} \cite{wang2017residual} which was proposed for image classification. The proposed MTFA includes a \fb{} for feature extraction and a \mb{} for attention generation.

We choose ResNet \cite{he2016deep} consisting of two residual blocks to construct the \fb{}. The kernel size of convolution within all residual blocks is set to \textit{3}$\times$\textit{3} because small kernel fuses less background with target events and is more sensitive to the onset. For the \mb{}, we utilize a bottom-up top-down structure called Hourglass \cite{newell2016stacked} which has shown its effectiveness in human pose estimation and image segmentation. Hourglass aims to fuse information from multiple resolutions, which aligns with our intention of multi-scale learning. While encoding coarse-grained information by downsampling feature maps to lower and lower resolutions along both time and frequency axis, Hourglass uses skip connection to maintain fine-grained information as well (the residual block within each skip connection is not shown in Figure \ref{fig:topology} for brevity). In this work we use three Max-pooling layers with kernel size and stride of
\textit{2}$\times$\textit{2} for downsampling, resulting in an observation at four different scales: $(T\times D), (T/2\times D/2), (T/4\times D/4), (T/8\times D/8)$, where $T$ and $D$ correspond to the length of MTFA's input along time and frequency axis, respectively. Then, low-resolution feature maps are upsampled back to the original scale using Nearest-Neighbor interpolation, and then merged with high-resolution representations which come through skip connections. Finally, the Sigmoid function is applied to generate an attention heatmap whose elements range within [0, 1].

Suppose the input of MTFA has a size of $(C\times T\times D)$, where $C$ denotes the number of feature channels within MTFA, and it is shared by the \fb{} and the \mb{}. Then, the output $M$ of the \mb{} has the same size as the representation $F$ produced by the \fb{}, which is also $(C\times T\times D)$. To get attention-aware features, one can simply perform element-wise multiplication between $M$ and $F$, i.e., $M\otimes F$. However, since $M$ ranges from 0 to 1, this naive production might degrade the values and make it difficult for the following propagation. Thus, following \cite{wang2017residual}, we use \xt{residual connection} \cite{he2016deep} and set the output $A$ of MTFA to be
\begin{equation}
A(x) = (1+M(x))\otimes F(x),
\end{equation}
where $x$ represents the input of MTFA module.


\subsection{Recurrent Neural Network}
We use a RNN to capture temporal dependencies between frames, which has been proved to work well when being combined with the CNN architecture. Here we use a bi-directional \xt{Gated Recurrent Unit} (GRU) which has $U$ hidden units and two layers, leading the size of the final high-level feature map to be $(T\times 2U)$. Fully-connected layers and sigmoid function are applied to output frame-wise predictions.

\begin{table*}[!t]
  \centering
  \caption{Performance comparision between the proposed method and previous ones. Event-based error rate and F1-score are evaluated following the setting of DCASE2017. *** indicates the corresponding results are not presented in the papers.}
  \label{tab:results}
  \begin{threeparttable}
    \begin{tabular}{lcccccccc}
    \toprule
    \multirow{2}{*}{Method}&
    \multicolumn{4}{c}{Development Dataset}&\multicolumn{4}{c}{Evaluation Dataset}\cr
    \cmidrule(lr){2-5} \cmidrule(lr){6-9}
    &babycry&glassbreak&gunshot&average&babycry&glassbreak&gunshot&average\cr
    &\scriptsize{ER$|$F1} &\scriptsize{ER$|$F1} &\scriptsize{ER$|$F1} &\scriptsize{ER$|$F1} &\scriptsize{ER$|$F1} &\scriptsize{ER$|$F1} &\scriptsize{ER$|$F1} &\scriptsize{ER$|$F1}\cr
    \midrule
    ResRNN &0.09$|$95.8 &0.03$|$98.4 &0.17$|$91.0 &0.10$|$95.1 &0.22$|$89.3 &0.07$|$96.4 &0.24$|$87.4 &0.18$|$91.0\cr
    ResAttRNN &0.13$|$93.4 &0.02$|$98.8 &0.18$|$90.4 &0.11$|$94.2 &0.22$|$88.9 &0.04$|$97.8 &0.17$|$91.2 &0.14$|$92.6\cr
    {\bf Proposed} &0.06$|$96.7 &0.02$|$99.0 &{\bf0.14$|$92.7}  &0.07$|$96.1 &{\bf0.10$|$95.1} &{\bf0.02$|$98.8} &{\bf0.14$|$92.5} &{\bf0.09$|$95.5}\\ \midrule
    1D-CRNN \cite{Lim2017}  &{\bf0.05$|$97.6} &{\bf0.01$|$99.6} &0.16$|$91.6 &{\bf0.07$|$96.3} &0.15$|$92.2 &0.05$|$97.6 &0.19$|$89.6 &0.13$|$93.1\cr
    CRNN \cite{Cakir2017}   &****** &****** &****** &0.14$|$92.9 &0.18$|$90.8 &0.10$|$94.7 &0.23$|$87.4 &0.17$|$91.0 \cr
    R-CRNN \cite{Kao2018}   &0.09$|$***~ &0.04$|$***~ &0.14$|$***~ &0.09$|$95.5 &****** &****** &****** &0.23$|$87.9\cr
    Multi-Scale RNN \cite{Wang2018} &0.11$|$94.3 &0.04$|$97.8 &0.18$|$90.6 &0.11$|$94.2 &0.26$|$86.5 &0.16$|$92.1 &0.18$|$91.1 &0.20$|$89.9\cr
    CRNN+Attention \cite{shen2018learning} &0.10$|$95.1 &0.01$|$99.4 &0.16$|$91.5 &0.09$|$95.3 &0.18$|$91.3 &0.04$|$98.2 &0.17$|$90.8 &0.13$|$93.4\cr
    \bottomrule
    \end{tabular}
  \end{threeparttable}
\end{table*}

\section{Experiments}

\subsection{Dataset}
We evaluate the proposed method on Task 2 of DCASE2017 Challenge \cite{DCASE2017challenge}. The dataset consists of 30-second sound clips, each of which has one isolated target event (either \babycry{}, \glassbreak{}, or \gunshot{}) along with recordings of different acoustic scenes taken from TUT Acoustic Scenes 2016 \cite{mesaros2016tut} as the background. We use the provided synthesizer to generate mixtures with different event-to-background ratios and random onset time. For the train set of development dataset, we generate 5000 mixtures for each target event with event-to-background ratios of -6, 0, 6dB and event presence probability of 0.99 in order to have more positive samples. For the test set of development dataset, 500 mixtures for each event are generated following the default parameters for fair comparison. The training data for development and evaluation dataset is the same.

\subsection{Experiment Setup}
During the training phase, the model takes a chunk of the spectrogram as input, i.e., several consecutive frames, rather than the whole spectrogram to increase batch size and training speed. In our experiments, the temporal length $T$ is set to 256 and the shift between each chunk is 128. During the inference phase, instead, we pass the spectrogram as a whole into the network. Since the input will be downsampled trice during propagation, we pad the spectrogram with the last frame to satisfy this length constraint. The number of channels within the MTFA module $C$ is set to 64, and the number of GRU units $U$ is also 64. We apply dropout \cite{srivastava2014dropout} with a rate of 0.3 for \babycry{} and \glassbreak{}, and 0.4 for \gunshot{} as we observe more severe overfitting with it.

The task could be regarded as a 0/1 classification since our method makes frame-wise predictions. Therefore, we use binary cross-entropy as the loss function. We use Adam \cite{kingma2014adam} as the optimizer with a learning rate of 0.001. The training is terminated when the validation loss has stopped improving for 10 epochs, and all experiments finished training within 40 epochs according to our observation. The model with the lowest validation loss is selected for the test. We implement our models with PyTorch \cite{paszke2017automatic}.

When the model is to be evaluated, the output occurrence probabilities are binarized with thresholds of 0.4, 0.2, and 0.4 for \babycry{}, \glassbreak{}, and \gunshot{}, respectively. The binary predictions are further processed with a 540ms median filter, which follows the default setting of DCASE2017 baseline.

\subsection{Evaluation Metrics}
Event-based error rate and F1-score are two metrics used for DCASE2017 Task2. They are calculated using onset-only condition with a collar of 500ms. The evaluation is done automatically using sed\_eval toolbox provided by the task organizers. Details can be found in \cite{Mesaros2016_MDPI}.

\begin{figure}[!t]
  \centering
  \includegraphics[width=3in]{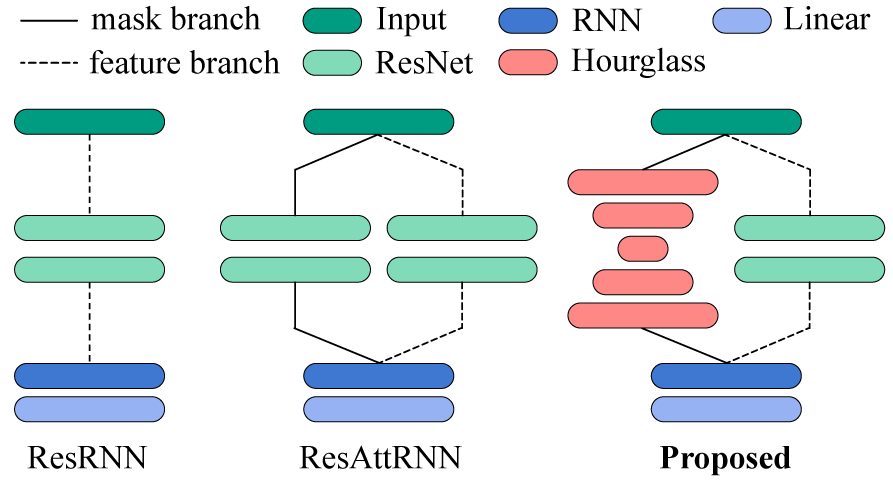}
  \caption{A simple overview of structures for \textbf{ablation study}.}
  \label{fig:ablation}
\end{figure}

\subsection{Results}
{\bf Ablation studies.} Two more models are studied here to identify the contribution of the proposed MTFA. We construct the first one by removing the \mb{} from MTFA and leaving only the \fb{}, which is termed ResRNN. The second one is built by replacing the Hourglass in the \mb{} with a ResNet, and we name it ResAttRNN. The general structures are shown in Figure \ref{fig:ablation}. In a nutshell, while ResRNN does not perform any attention techniques, ResAttRNN generates temporal-frequential attention at only one single scale. Results are presented in the first three rows of Table \ref{tab:results}. While ResRNN reaches comparable results to those of ResAttRNN on development dataset, its performance degrades a lot on evaluation dataset and is worse than ResAttRNN, indicating that temporal-frequential attention indeed helps to extract powerful features which could generalize better. Furthermore, compared with ResAttRNN, MTFA performs better on both development and evaluation dataset, confirming that multi-scale learning can further improve the performance.

{\bf Comparison with previous methods.} We present results in Table \ref{tab:results}. On development dataset, the proposed method achieves equally low error rate as R-CRNN \cite{Kao2018} on \gunshot{} and also gets desirable results on \babycry{} and \glassbreak{}. Note that 1D-CRNN \cite{Lim2017} adopted ensemble strategy and fused the output of up to 5 models to get final predictions. However, our method could perform comparably well without ensemble and achieves the best average result compared with other single-model methods like \cite{Kao2018, Wang2018} and especially \cite{shen2018learning} which also generated temporal-frequential attention but in a different manner. Note that for all methods, the performance on \gunshot{} is the worst because the sound varies with different gun types and presents different temporal and frequential properties, which supports our analysis in the beginning.

On evaluation dataset, our method outperforms all previous methods on each target event. With the average error rate slightly increases from 0.07 on development dataset to 0.09 on evaluation dataset, it is confirmed that MTFA has a better generalization ability. To sum up, the above results demonstrate that with the help of multi-scale learning and attention mechanism, MTFA is more powerful an option than previous methods \cite{Lim2017, Cakir2017, Kao2018, Wang2018, shen2018learning} to achieve discriminative features.

\begin{figure}[!t]
  \centering
  \includegraphics[width=3in]{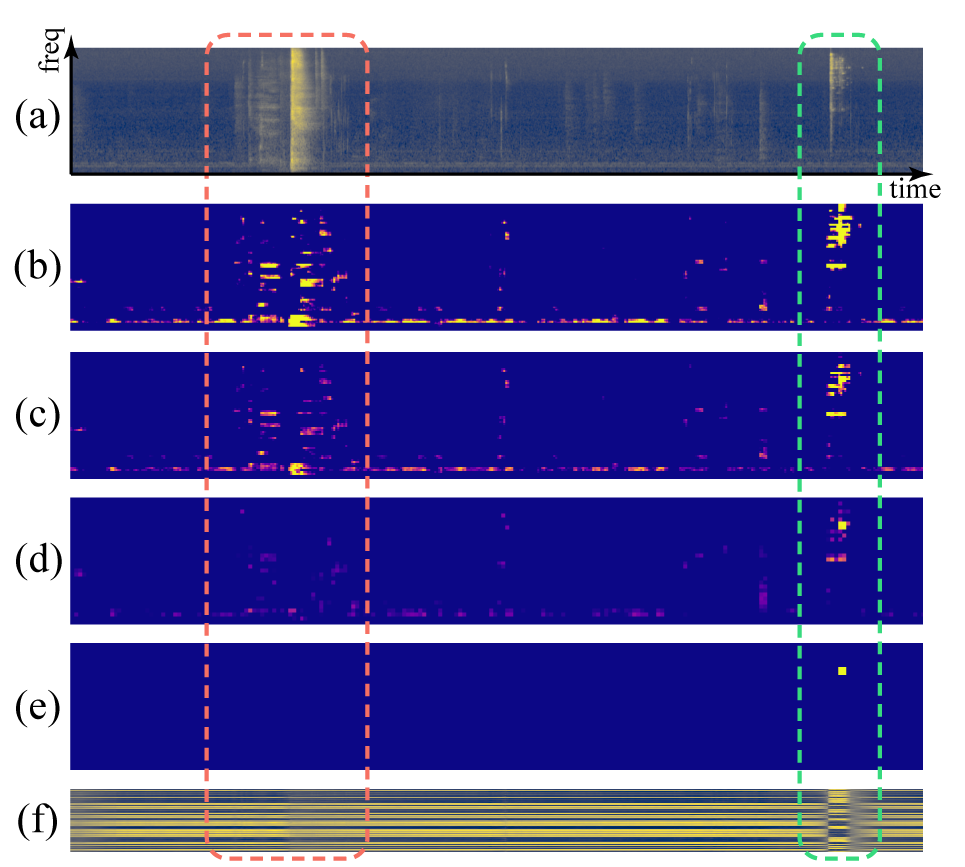}
  \caption{\textbf{Visualization of MTFA.} Taking glassbreak as an example, we visualize (a) the input spectrogram and (f) the output of RNN. From (b) to (e) are the outputs of MTFA from four scales with resolution from high to low. The target glassbreak is enclosed by the green box, while another sound which might interfere the prediction is enclosed by the pink box.}
  \label{fig:visualization}
\end{figure}

\subsection{Visualization of MTFA}
We visualize the output of the MTFA module in Figure \ref{fig:visualization} to provide a straightforward and intuitive understanding. A sound clip from evaluation dataset is taken as an example. As shown by the spectrogram \textit{(a)}, besides the target \glassbreak{} near the end (enclosed by the green box), the audio clip also comprises a man's `cough' at the beginning (enclosed by the pink box). From \textit{(b)} to \textit{(e)}, we present the outputs of the MTFA's \mb{} extracted from four scales, where \textit{(b)} and \textit{(e)} has the highest and lowest resolution, respectively. It is obvious that low-resolution representations like \textit{(d)} and \textit{(e)} will drop details and only leave high-level information which only considers the location of the target event. Discarding details plays an important role here, since this move arms the model with the robustness against temporal and frequential variations, i.e., different scales possessed by events won't affect the high-level low-resolution results presented by \textit{(d)} and \textit{(e)}. On the contrary, if analysis is performed using single resolution, like only with \textit{(b)}, the model might be more vulnerable to scale variations and produce worse performance, just like ResRNN and ResAttRNN in Table \ref{tab:results}.

Meanwhile, despite that \textit{(b)} assigns higher weights to background noise and the `cough' sound beside \glassbreak{}, \textit{(c)}, \textit{(d)}, and \textit{(e)} gradually suppress these interferences more and more and only pay attention to the target event, making it quite easy for the following RNN to identify the target event, as clearly shown in \textit{(f)}. The visualization results provide another evidence that our MTFA is an effective approach for AED.

\section{Conclusion}

In this work, we propose a \xt{Multi-Scale Time-Frequency Attention} module for acoustic event detection. Multi-scale learning is incorporated to capture the properties of events with different temporal and frequential scales, and time-frequency attention is applied to instruct the model to focus on not only important frames but also important frequency components. We evaluate the proposed method on DCASE2017 Task2. It turns out that MTFA is the best-performing single-model method on development dataset and achieves state-of-the-art on evaluation dataset, which proves its effectiveness on extracting discriminative multi-scale attention-aware features. We hope this work can provide insight for researchers on adopting multi-scale learning and temporal-frequential attention for acoustic-related tasks.

\section{Acknowledgement}
This work was supported by the National Natural Science Foundation of China under Grant No. 61403224 and No. U1836219.

\bibliographystyle{IEEEtran}

\bibliography{mybib}

\begin{thebibliography}{10}
\providecommand{\url}[1]{#1}
\csname url@samestyle\endcsname
\providecommand{\newblock}{\relax}
\providecommand{\bibinfo}[2]{#2}
\providecommand{\BIBentrySTDinterwordspacing}{\spaceskip=0pt\relax}
\providecommand{\BIBentryALTinterwordstretchfactor}{4}
\providecommand{\BIBentryALTinterwordspacing}{\spaceskip=\fontdimen2\font plus
\BIBentryALTinterwordstretchfactor\fontdimen3\font minus
  \fontdimen4\font\relax}
\providecommand{\BIBforeignlanguage}[2]{{%
\expandafter\ifx\csname l@#1\endcsname\relax
\typeout{** WARNING: IEEEtran.bst: No hyphenation pattern has been}%
\typeout{** loaded for the language `#1'. Using the pattern for}%
\typeout{** the default language instead.}%
\else
\language=\csname l@#1\endcsname
\fi
#2}}
\providecommand{\BIBdecl}{\relax}
\BIBdecl

\bibitem{schroder2013automatic}
J.~Schr{\"o}der, S.~Goetze, V.~Gr{\"u}tzmacher, and J.~Anem{\"u}ller,
  ``Automatic acoustic siren detection in traffic noise by part-based models,''
  in \emph{2013 IEEE International Conference on Acoustics, Speech and Signal
  Processing}.\hskip 1em plus 0.5em minus 0.4em\relax IEEE, 2013, pp. 493--497.

\bibitem{wu2009intelligent}
X.~Wu, H.~Gong, P.~Chen, Z.~Zhi, and Y.~Xu, ``Intelligent household
  surveillance robot,'' in \emph{2008 IEEE International Conference on Robotics
  and Biomimetics}.\hskip 1em plus 0.5em minus 0.4em\relax IEEE, 2009, pp.
  1734--1739.

\bibitem{foggia2016audio}
P.~Foggia, N.~Petkov, A.~Saggese, N.~Strisciuglio, and M.~Vento, ``Audio
  surveillance of roads: A system for detecting anomalous sounds,'' \emph{IEEE
  transactions on intelligent transportation systems}, vol.~17, no.~1, pp.
  279--288, 2016.

\bibitem{DCASE2017challenge}
A.~Mesaros, T.~Heittola, A.~Diment, B.~Elizalde, A.~Shah, E.~Vincent, B.~Raj,
  and T.~Virtanen, ``{DCASE} 2017 challenge setup: Tasks, datasets and baseline
  system,'' in \emph{Proceedings of the Detection and Classification of
  Acoustic Scenes and Events 2017 Workshop (DCASE2017)}, November 2017, pp.
  85--92.

\bibitem{Lim2017}
H.~Lim, J.~Park, and Y.~Han, ``Rare sound event detection using {1D}
  convolutional recurrent neural networks,'' DCASE2017 Challenge, Tech. Rep.,
  September 2017.

\bibitem{Cakir2017}
E.~Cakir and T.~Virtanen, ``Convolutional recurrent neural networks for rare
  sound event detection,'' DCASE2017 Challenge, Tech. Rep., September 2017.

\bibitem{Kao2018}
\BIBentryALTinterwordspacing
C.-C. Kao, W.~Wang, M.~Sun, and C.~Wang, ``R-crnn: Region-based convolutional
  recurrent neural network for audio event detection,'' in \emph{Proc.
  Interspeech 2018}, 2018, pp. 1358--1362. [Online]. Available:
  \url{http://dx.doi.org/10.21437/Interspeech.2018-2323}
\BIBentrySTDinterwordspacing

\bibitem{ren2015faster}
S.~Ren, K.~He, R.~Girshick, and J.~Sun, ``Faster r-cnn: Towards real-time
  object detection with region proposal networks,'' in \emph{Advances in neural
  information processing systems}, 2015, pp. 91--99.

\bibitem{Wang2018}
\BIBentryALTinterwordspacing
W.~Wang, C.-C. Kao, and C.~Wang, ``A simple model for detection of rare sound
  events,'' in \emph{Proc. Interspeech 2018}, 2018, pp. 1344--1348. [Online].
  Available: \url{http://dx.doi.org/10.21437/Interspeech.2018-2338}
\BIBentrySTDinterwordspacing

\bibitem{Luo2018}
\BIBentryALTinterwordspacing
D.~Luo, Y.~Zou, and D.~Huang, ``Investigation on joint representation learning
  for robust feature extraction in speech emotion recognition,'' in \emph{Proc.
  Interspeech 2018}, 2018, pp. 152--156. [Online]. Available:
  \url{http://dx.doi.org/10.21437/Interspeech.2018-1832}
\BIBentrySTDinterwordspacing

\bibitem{Bothe2018}
\BIBentryALTinterwordspacing
C.~Bothe, S.~Magg, C.~Weber, and S.~Wermter, ``Conversational analysis using
  utterance-level attention-based bidirectional recurrent neural networks,'' in
  \emph{Proc. Interspeech 2018}, 2018, pp. 996--1000. [Online]. Available:
  \url{http://dx.doi.org/10.21437/Interspeech.2018-2527}
\BIBentrySTDinterwordspacing

\bibitem{Chang2018}
\BIBentryALTinterwordspacing
X.~Chang, Y.~Qian, and D.~Yu, ``Monaural multi-talker speech recognition with
  attention mechanism and gated convolutional networks,'' in \emph{Proc.
  Interspeech 2018}, 2018, pp. 1586--1590. [Online]. Available:
  \url{http://dx.doi.org/10.21437/Interspeech.2018-1547}
\BIBentrySTDinterwordspacing

\bibitem{wang2017residual}
F.~Wang, M.~Jiang, C.~Qian, S.~Yang, C.~Li, H.~Zhang, X.~Wang, and X.~Tang,
  ``Residual attention network for image classification,'' in \emph{Proceedings
  of the IEEE Conference on Computer Vision and Pattern Recognition}, 2017, pp.
  3156--3164.

\bibitem{shen2018learning}
Y.-H. Shen, K.-X. He, and W.-Q. Zhang, ``Learning how to listen: A
  temporal-frequential attention model for sound event detection,'' \emph{arXiv
  preprint arXiv:1810.11939}, 2018.

\bibitem{he2016deep}
K.~He, X.~Zhang, S.~Ren, and J.~Sun, ``Deep residual learning for image
  recognition,'' in \emph{Proceedings of the IEEE conference on computer vision
  and pattern recognition}, 2016, pp. 770--778.

\bibitem{newell2016stacked}
A.~Newell, K.~Yang, and J.~Deng, ``Stacked hourglass networks for human pose
  estimation,'' in \emph{European Conference on Computer Vision}.\hskip 1em
  plus 0.5em minus 0.4em\relax Springer, 2016, pp. 483--499.

\bibitem{mesaros2016tut}
A.~Mesaros, T.~Heittola, and T.~Virtanen, ``Tut database for acoustic scene
  classification and sound event detection,'' in \emph{2016 24th European
  Signal Processing Conference (EUSIPCO)}.\hskip 1em plus 0.5em minus
  0.4em\relax IEEE, 2016, pp. 1128--1132.

\bibitem{srivastava2014dropout}
N.~Srivastava, G.~Hinton, A.~Krizhevsky, I.~Sutskever, and R.~Salakhutdinov,
  ``Dropout: a simple way to prevent neural networks from overfitting,''
  \emph{The Journal of Machine Learning Research}, vol.~15, no.~1, pp.
  1929--1958, 2014.

\bibitem{kingma2014adam}
D.~P. Kingma and J.~Ba, ``Adam: A method for stochastic optimization,''
  \emph{arXiv preprint arXiv:1412.6980}, 2014.

\bibitem{paszke2017automatic}
A.~Paszke, S.~Gross, S.~Chintala, G.~Chanan, E.~Yang, Z.~DeVito, Z.~Lin,
  A.~Desmaison, L.~Antiga, and A.~Lerer, ``Automatic differentiation in
  pytorch,'' in \emph{NIPS-W}, 2017.

\bibitem{Mesaros2016_MDPI}
\BIBentryALTinterwordspacing
A.~Mesaros, T.~Heittola, and T.~Virtanen, ``Metrics for polyphonic sound event
  detection,'' \emph{Applied Sciences}, vol.~6, no.~6, p. 162, 2016. [Online].
  Available: \url{http://www.mdpi.com/2076-3417/6/6/162}
\BIBentrySTDinterwordspacing

\end{thebibliography}

\end{document}